\documentclass{PoS}

\usepackage{epsfig}
\usepackage{amsmath}
\usepackage{amssymb}

\PoS{PoS(LAT2005)161}

\title{Searching for the critical endpoint in QCD with two quark flavors }

\ShortTitle{Searching for the critical endpoint in QCD with two quark 
flavors }

\author{\speaker{Victor Laliena}\thanks{Ram\'on y Cajal Fellow}\\
        Universidad de Zaragoza\\
        E-mail: \email{laliena@unizar.es}}

\abstract{I present a method which can be used to locate the expected
critical endpoint of the phase diagram of QCD with two light quark flavors.
I illustrate the ideas on a suitable Random Matrix Model and  
show preliminary results in QCD}

\FullConference{XXIIIrd International Symposium on Lattice Field Theory\\
		 25-30 July 2005\\
		 Trinity College, Dublin, Ireland}

\begin{document}

\section{The expected phase diagram of QCD}

It is belived that the phase diagram of QCD is very rich, especially at 
low temperatures, where many different phases separated 
by phase boundaries  are expected. The analysis which lead to these 
expectations 
are based on symmetries, effective theories, models, and analogies, and
therefore are far from being well established \cite{kogut:book}. 

The major problem in the study
of the QCD phase diagram is that standard Monte Carlo simulations 
cannot be used in QCD with baryon chemical
potential since the fermion determinant is complex and cannot be included
in a probability measure. It is the so called sign problem.
But even at zero chemical potential the situation is not clear: in QCD with
two degenerate light quarks the nature of the deconfinement transition at 
finite temperature is not definitely established, and it is not even clear 
whether
it is a crossover or a true phase transition, either of first or second order 
class \cite{digiacomo}.
Until recently, it has been generally accepted that it is a crossover,
in which case the change from the confinement to the deconfinement regime
at high temperature and low baryon chemical potential would also be a
crossover.
Since a first order deconfinement transition is 
expected at sufficiently low temperatures, there must be a critical point
located where the first order transition ceases to exist and is replaced
by a crossover.
This is the critical endpoint. The expected phase diagram is depicted in
Figure~1. A Chiral Random Matrix Model predicts this phase structure
\cite{rmm}. Lattice computations overcoming the sign problem by means of
the double
reweighting technique have found signals of a critical point \cite{fodor}.

In the following we will propose a method to locate the critical endpoint
and present some preliminary results.
We assume that at zero chemical potential the confinement and deconfinement 
regimes are separated by a crossover.

\begin{figure}[b]
\centerline{\epsfig{file=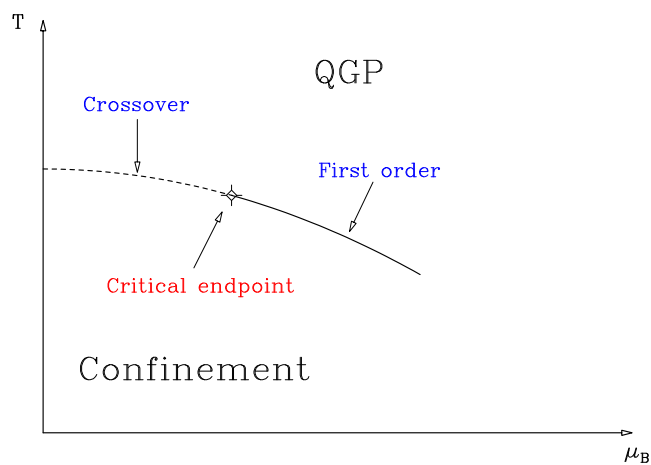,width=.4\textwidth,angle=90}}
\caption{The expected phase diagram of two flavor QCD}
\label{fig1}
\end{figure}

\section{The method}

Let us describe the general setting. A more complete discussion can be 
found  elsewhere \cite{nos}.
We work on a lattice with a generalized staggered fermion action
\begin{equation}
S = \frac{1}{2}\sum_n\sum^3_{i=1}\bar\psi_n \eta_i (n)
\left( U_{n,i} \psi_{n+i} - U^\dagger_{n-i,i}\psi_{n-i}\right)
+m\sum_n\bar\psi_n\psi_n + S_\tau\, ,
\end{equation}
where $S_\tau$ contains the temporal (covariant) derivative, and therefore
the temporal links:
\begin{equation}
S_{\tau} \;=\;\frac{x}{2} \sum_{n}\bar\psi_n \eta_0 (n)\left( U_{n,0}
\psi_{n+0} - U^\dagger_{n-0,0}\psi_{n-0}\right) 
\;+\; \frac{y}{2} \sum_{n} \bar\psi_n \eta_0 (n)\left(  U_{n,0}
\psi_{n+0} + U^\dagger_{n-0,0}\psi_{n-0}\right)\, .
\end{equation}
We consider $x$ and $y$ as independent parameters. QCD is recovered by
setting $x=\cosh\mu$ and $y=\sinh\mu$, where $\mu$ is the baryon
chemical potential. The sign problem is present if $\mathrm{Re}\,y\neq0$. 
However, it is easy to see that there is no sign problem if $y$ is purely
imaginary, $y=\mathrm{i}\bar{y}$, with $\bar{y}$ real. Imaginary chemical 
potential corresponds to $x=\cos\mu$ and $y=\mathrm{i}\sin\mu$.

It can be seen that the partition function (and therefore the phase diagram)
depends on $x$ and $y$ only through the following two variables:
\begin{equation}
u \;=\; x^2-y^2\, , \hspace{1truecm}
v \;=\; (x+y)^{3L_t}+(x-y)^{3L_t}\, ,
\end{equation}
where $L_t$ is the lattice size in the temporal direction. This two variables
are real for both real and imaginary $y$. Also, they are invariant under the
change $y\rightarrow -y$ (or $\bar{y}\rightarrow -\bar{y}$) and
$x\rightarrow -x$ separately, since $L_t$ is even. Altogether, this implies
that the phase diagram can be plotted in the $(x^2,y^2)$ plane. Obviously,
the semiplane $y^2<0$ corresponds to imaginary $y$. The physical region is 
the line
$x^2-y^2=1$, with $x^2\geq 1$. Imaginary chemical potential corresponds to
the line $x^2-y^2=1$, with $0\leq x^2\leq 1$. Numerical simulations are 
feasible for $x^2\geq 0$ and $y^2\leq 0$.

The expected phase diagram projected onto the $(x^2,y^2)$ plane is
displayed in Figure~2. The extended phase diagram is three dimensional, with
the temperature axis, which is not displayed, perpendicular to the $x^2$
and $y^2$ axis. 
We assume that on the physical line (the black line
on the figure, made up of solid and dashed pieces), for large
$x$ (i.e, large chemical potential) there is a first order phase transition
(solid line), which ends at the critical endpoint, marked with an open symbol
on the figure. For lower $x$ on the physical line there is a crossover
(dashed line) separating the low temperature confining regime from the high
temperature deconfined phase. The first order line is extended over a surface
in the three dimensional phase diagram, and the critical point is extended 
along a line of critical points (blue line) which is the boundary of the 
first order surface (projected onto the $(x^2,y^2)$ plane on the figure).

\begin{figure}[b]
\centerline{\epsfig{file=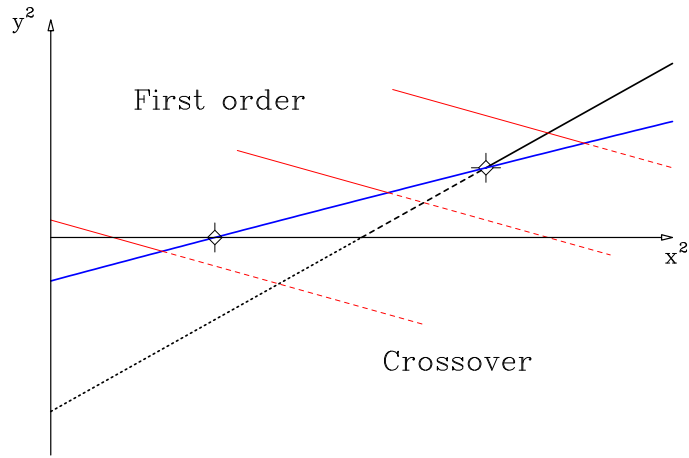,width=.4\textwidth,angle=90}}
\caption{The expected phase diagram projected onto the $(x^2,y^2)$ plane.
Although the transition lines are plotted as straight lines for simplicity,
it should not be inferred we expect they have such simple form.}
\label{fig2}
\end{figure}

The $y^2=0$ axis is interesting because numerical simulations are feasible 
there. At $x=1$ and $y=0$, which is the zero chemical potential point on the
physical line, there is a crossover, by assumption. On the other hand, at
$x=0$ and $y=0$ the fermion determinant contains no temporal link 
--it contains only the spatial links-- and
obviously cannot break the Polyakov symmetry explicitely. Hence, the Polyakov
loop is an order parameter. We expect center symmetry spontaneously broken at
high temperature and a first order transition separating the low and high
temperature phases. First order transitions are generically robust under
perturbations and therefore there must be a first order transition for $y^2=0$
and any $x^2$ sufficiently small. This first order line must end at a
critical point located at $x^2<1$ (marked with an open symbol on the figure), 
since we assume a crossover at $x^2=1$.
The natural scenario is a critical line crossing the physical line at
the physical critical endpoint and the $y^2=0$ axis at some point of the
segment $0<x^2<1$, as the blue line of Figure~2.

The critical line can be determined for $y^2\leq 0$ by means of standard 
Monte Carlo simulations and extended to small positive $y^2$ by double
reweighting. The resulting line might be extrapolated to the physical
line and then the position of the physical critical endpoint would have been 
determined. Even if the extrapolation were not reliable, finding a critical
line directed towards the physical line would be very interesting, since it
would indicate the existence of the critical endpoint.

\section{A Chiral Random Matrix Model}

A Chiral Random Matrix Model of QCD predics the existence of a critical
endpoint in the $(\mu,T)$ plane for light quarks \cite{rmm}. 
We can adapt this model
to study the structure of the extended phase diagram, in the $(x^2,y^2,T)$
space. To this end, let us consider the following partition function
\begin{equation}
\mathcal{Z}=\int [dA^\dagger dA][d\bar\psi_Ld\psi_Ld\bar\psi_Rd\psi_R]
\exp\left(-S\right)\, ,
\end{equation}
with
\begin{eqnarray}
S = N\sum_{t=0}^{L_t-1} \mathrm{Tr} (A^\dagger_t A_t)
-\sum_{t=0}^{L_t-1}\left[\bar\psi^a_L(t)\mathrm{i}A_t\psi^a_L(t)
+\bar\psi^a_R(t)\mathrm{i}A^\dagger_t\psi^a_R(t)\right] 
- m\sum_{t=0}^{L_t-1}\left[\bar\psi^a_L(t)\psi^a_R(t)
+\bar\psi^a_R(t)\psi^a_L(t)\right] \nonumber \\
-\beta \frac{x+y}{2}
\sum_{t=0}^{L_t-1}\left[\bar\psi^a_L(t)\psi^a_L(t+1)
+\bar\psi^a_R(t)\psi^a_R(t+1)\right] 
-\beta \frac{x-y}{2}
\sum_{t=0}^{L_t-1}\left[\bar\psi^a_L(t+1)\psi^a_L(t)
+\bar\psi^a_R(t+1)\psi^a_R(t)\right]\, , \nonumber
\end{eqnarray}
where $A_t$ is a complex $N\times N$ random matrix depending on the temporal 
index $t$ ($N$ collects the color, Dirac, and spatial indices),
$\psi^a_L(t)$ and $\psi^a_R(t)$ are respectively the left and right handed
components of a fermion field, and $a$ is a flavor index. The parameter $m$ is
the fermion mass, which will be set to zero in the following, $x$ and $y$ are 
the parameters introduced in the previous section, and $\beta$ controls the
temperature (large $\beta$ corresponds to high temperature). Notice that, in
order to introduce the chemical potential in the lattice way, we keep the
dynamics local (and free for simplicity) in the temporal direction.

The model can be solved analytically in the large $N$ limit with the standard
techniques. Figure~3 (left panel) displays the physical phase diagram 
($x=\cosh\mu$, $y=\sinh\mu$) for $N_t=4$. There are two phases: at low
temperature chiral symmetry is spontaneously broken and it is restored at
high temperatures. The transition is second order at small $\mu$ and first
order at large $\mu$. The separation between both lines is a tricritical
point. With a small mass, the second order line becomes a crossover, the 
first order transition remains and the tricritical point becomes the critical
endpoint. The phase diagram mimics perfectly what is expected in QCD.
Its extension to the $(x^2,y^2)$ plane, displayed on the right panel of
Figure~3, complies with our expectations: the black line is the physical line,
the solid and dashed red lines are respectively first and second order 
transitions, and the blue line is the tricritical line, which in this case 
happens to be straight. It crosses the $y^2=0$ axis at $x^2=y^2=0$.
This is a pathology of the model which can be easily understood.
Moreover, the $(x^2=0,y^2=0)$ point is singular and the tricritical line
could not be continued from it by, say, double reweighting. Notice however 
that this model is based only on chiral symmetry and does not incorporate
the dynamics of the Polyakov loop, which is the basis of the argument of 
the previous section. Hence, it is interesting that its extended phase diagram
showed the features we predicted.

\begin{figure}[t]
\begin{minipage}[h]{7truecm}
\centerline{\epsfig{file=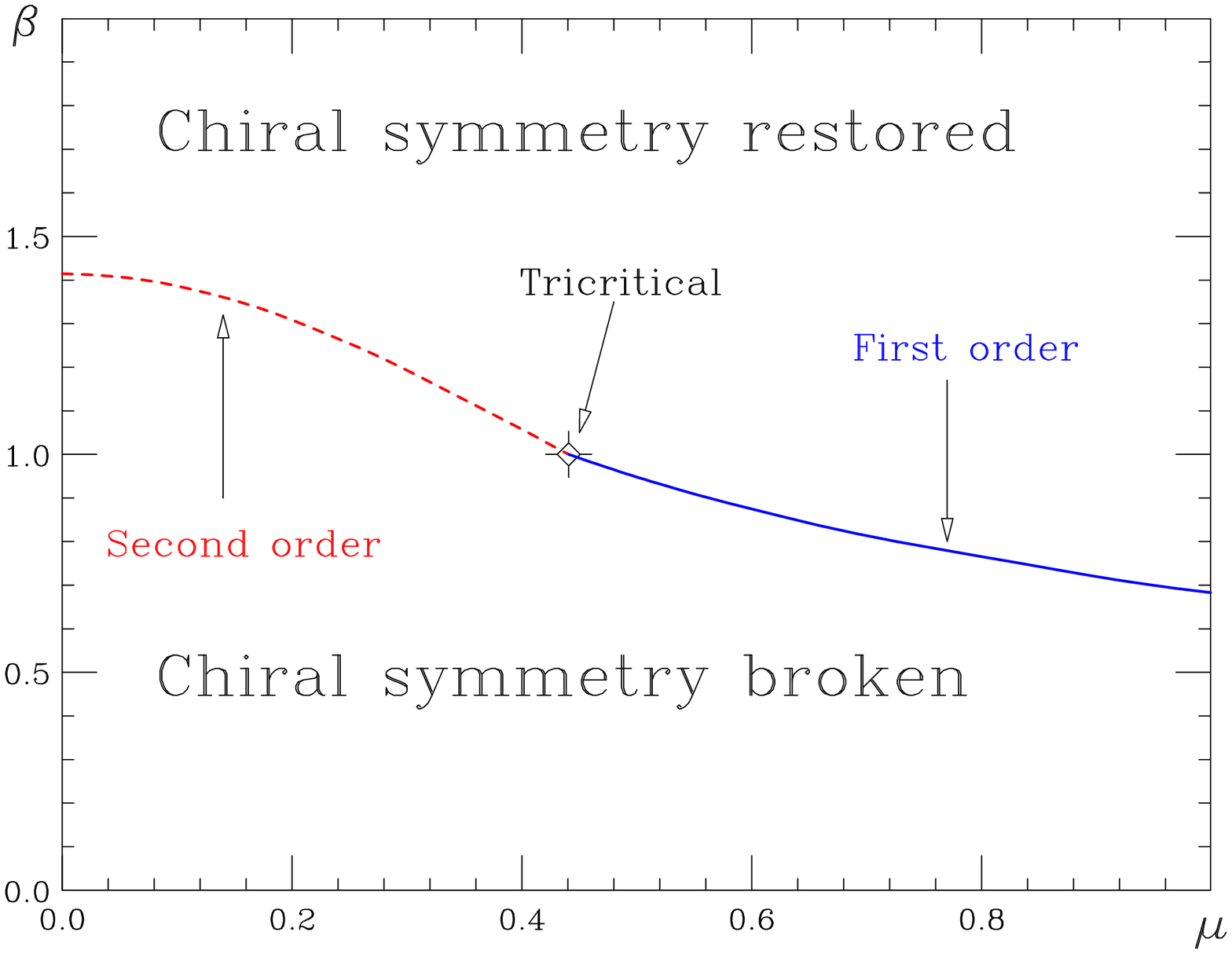,width=.7\textwidth,angle=90}}
\end{minipage}
\begin{minipage}[h]{7truecm}
\centerline{\epsfig{file=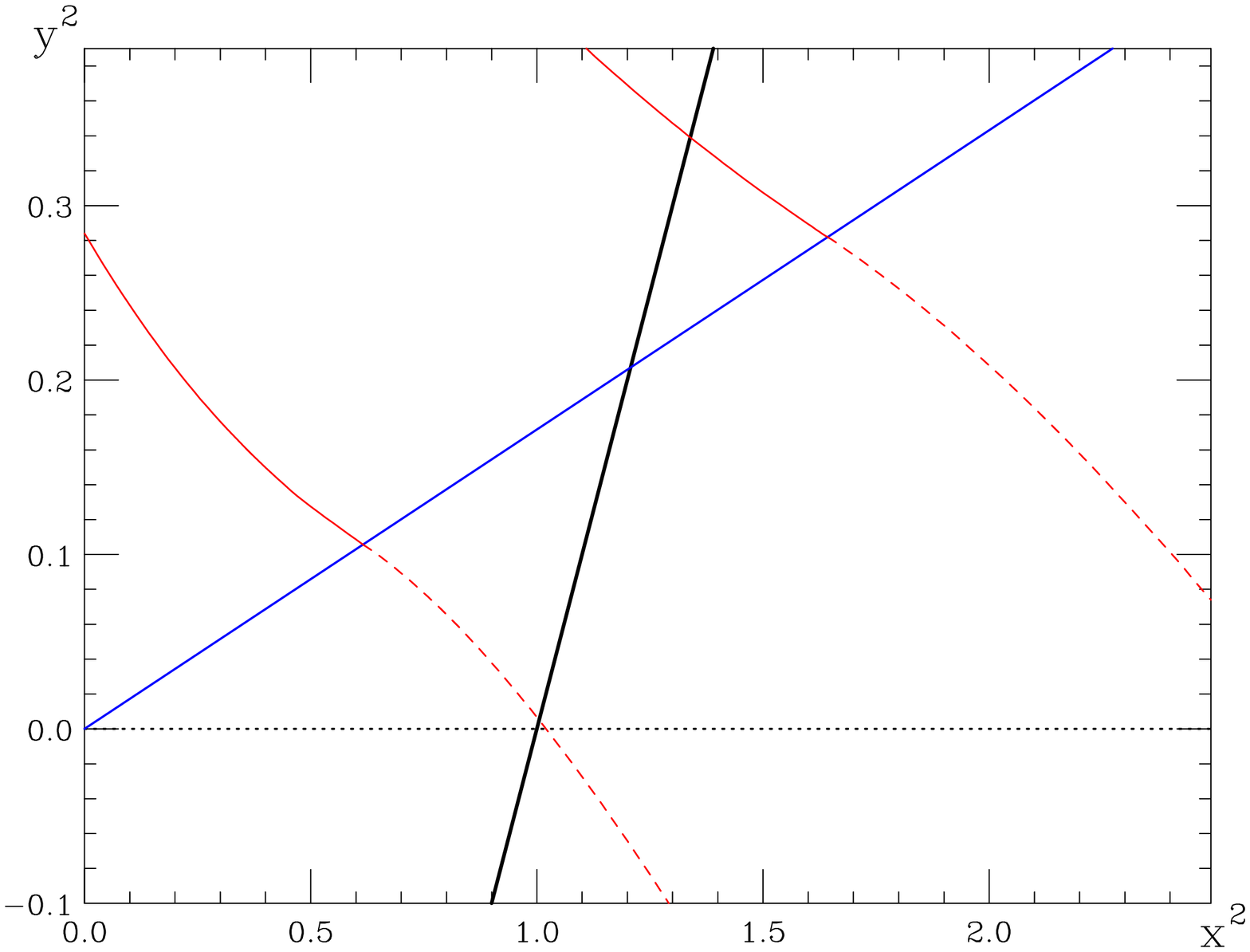,width=.7\textwidth,angle=90}}
\end{minipage}
\caption{The physical phase diagram of the CRMM (left) and its extension
to the $(x^2,y^2)$ plane (right).}
\label{fig3}
\end{figure}

\section{Preliminary results for QCD}

Figure~4 displays histograms of the modulus of the Polyakov loop
from simulations of QCD performed with the R-algorithm. Since it has
been pointed out that the systematics of this algorithm lead to 
wrong results for the time steps which use to be used \cite{kogut:sinclair}, 
the results should be taken with some caution (the extrapolation to zero 
time step should be addressed before extracting any conclusion). 
Anyway, they are very preliminary. 
We take two flavors of staggered quarks, with mass 
$ma=0.05$, and explore the $y=0$ axis on lattices of temporal extent
$L_t=4$ and spatial extent $L_s=6,8,10$. For each $x$ the value of the 
gauge coupling, $\beta$, has been tuned to its ``critical'' value.

For $x=0$ we see a kind of (not very clear) double pick structure on the 
$L_s=6$ lattice, wich becomes clearer as $L_s$ increases. The data point
out to a first order transition (remenber the action has exact Polyakov 
symmetry in this case). The situation for $x=0.5$ is similar.
For $x=0.7$, however, it is the opposite. We see a clear two pick structure
on the $L_s=6$ lattice which is disappearing as $L_s$ increases, signaling
a crossover.
There must be a critical point between $x=0.5$ and $x=0.7$.
Locating it accurately is a hard task which is in progress.

\begin{figure}[t]
\begin{minipage}[h]{7truecm}
\centerline{\epsfig{file=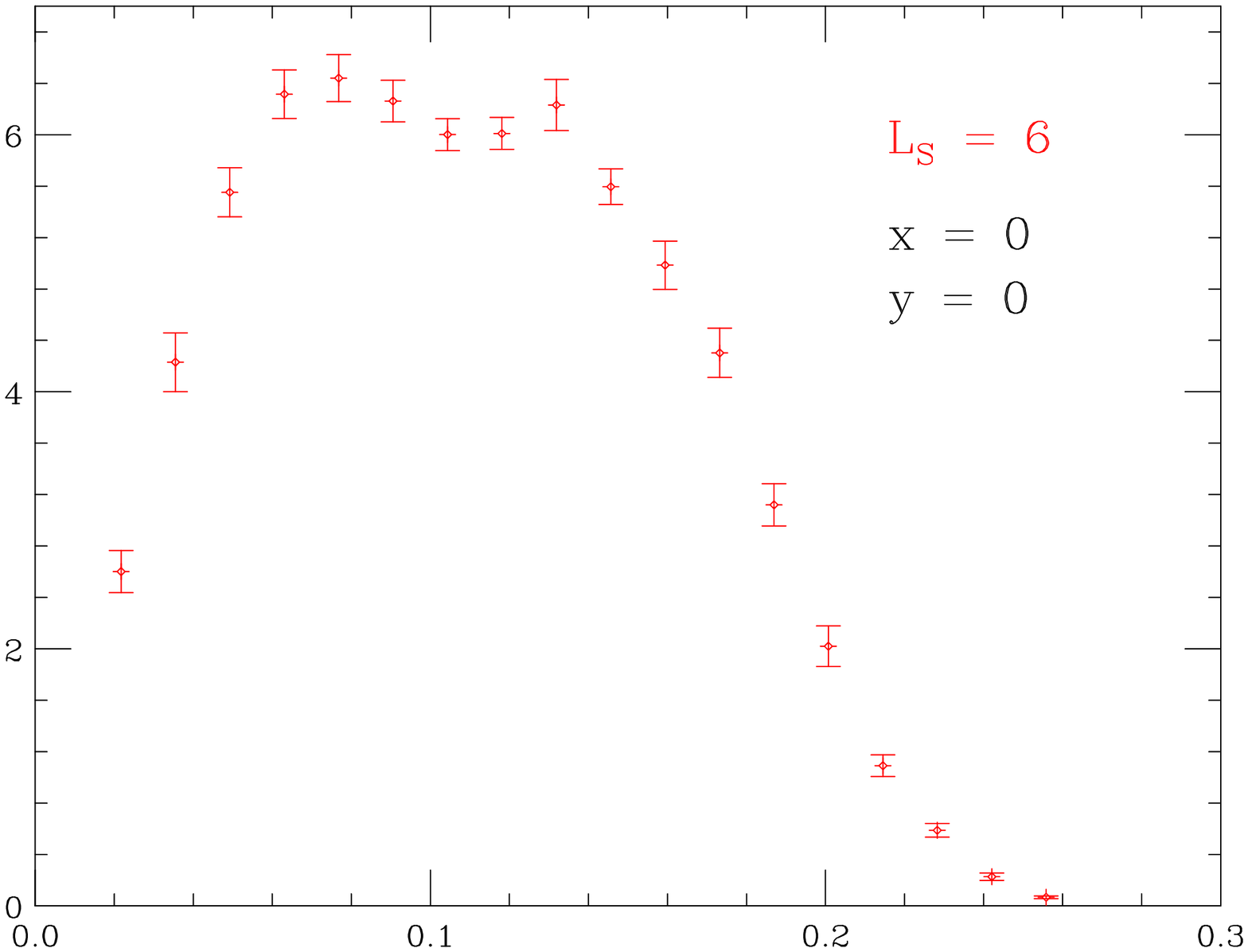,width=.7\textwidth,angle=90}}
\centerline{\epsfig{file=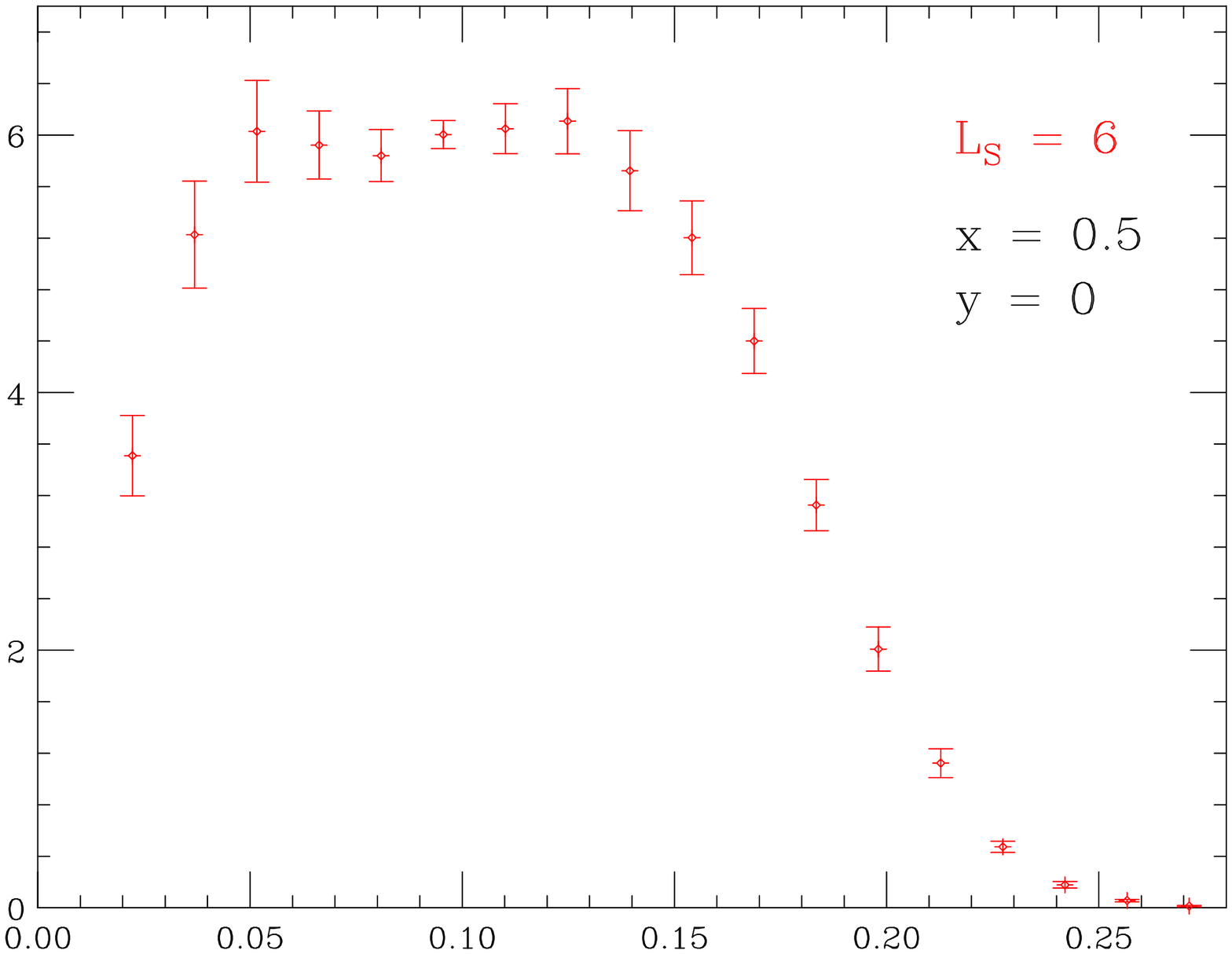,width=.7\textwidth,angle=90}}
\centerline{\epsfig{file=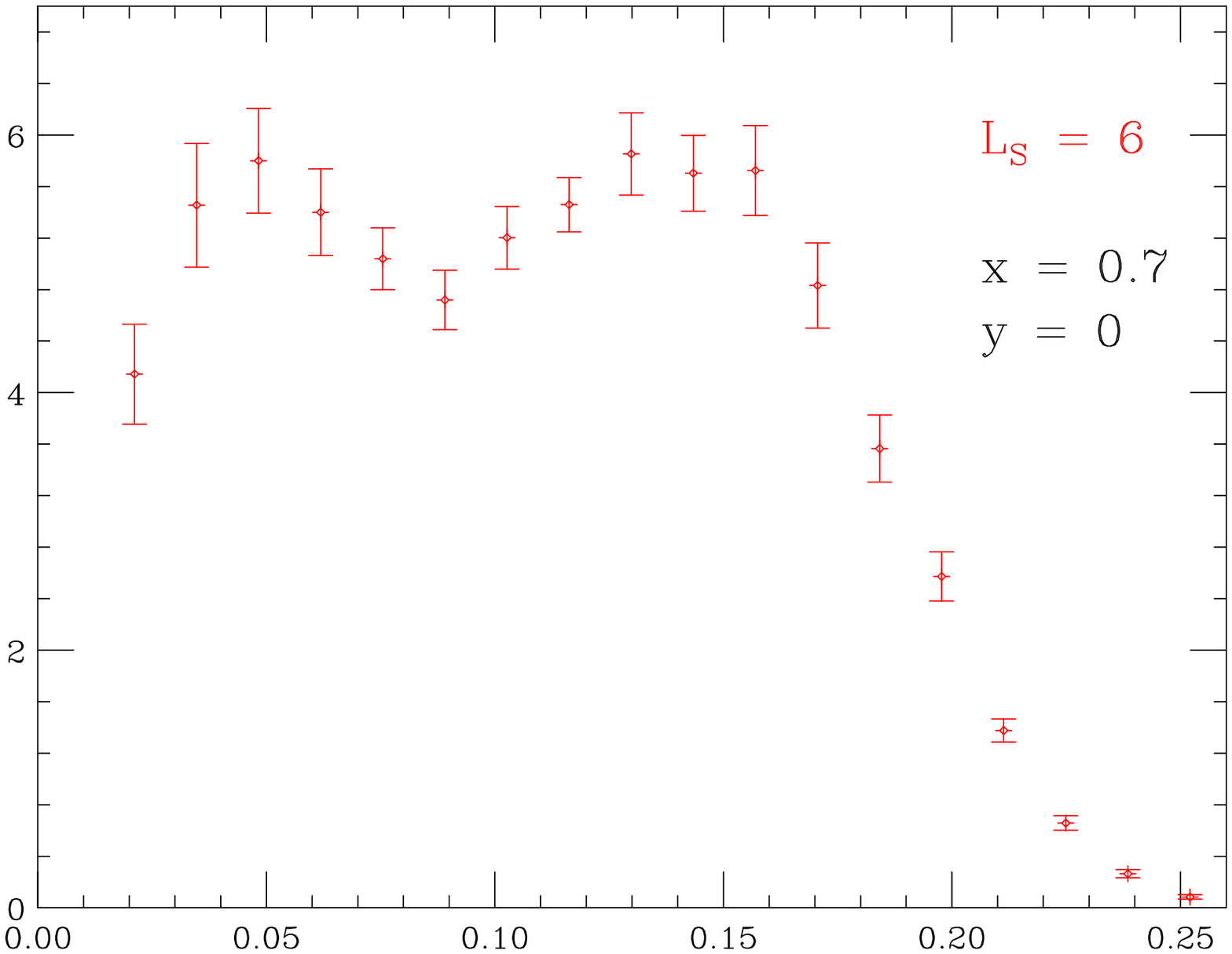,width=.7\textwidth,angle=90}}
\end{minipage}
\begin{minipage}[h]{7truecm}
\centerline{\epsfig{file=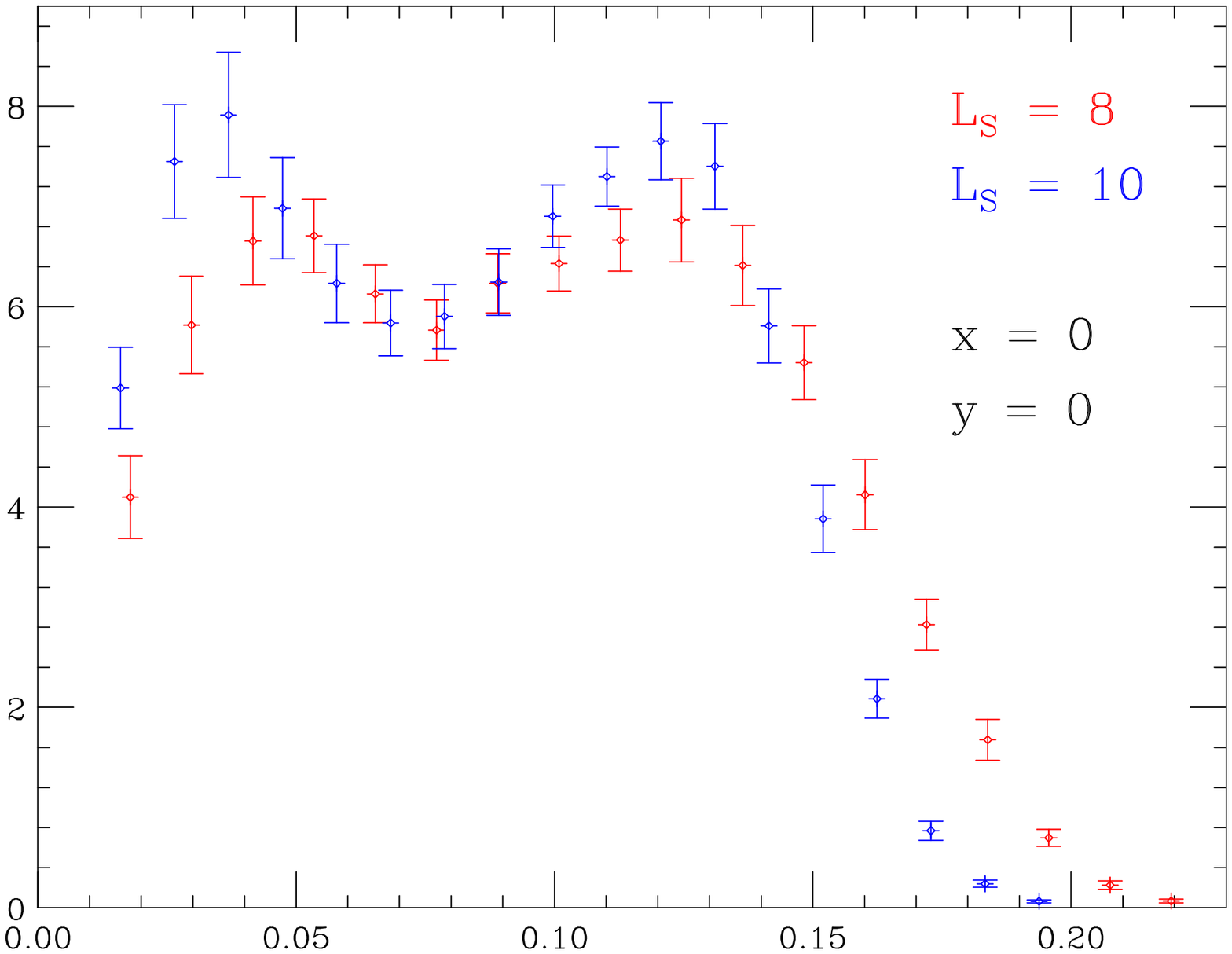,width=.7\textwidth,angle=90}}
\centerline{\epsfig{file=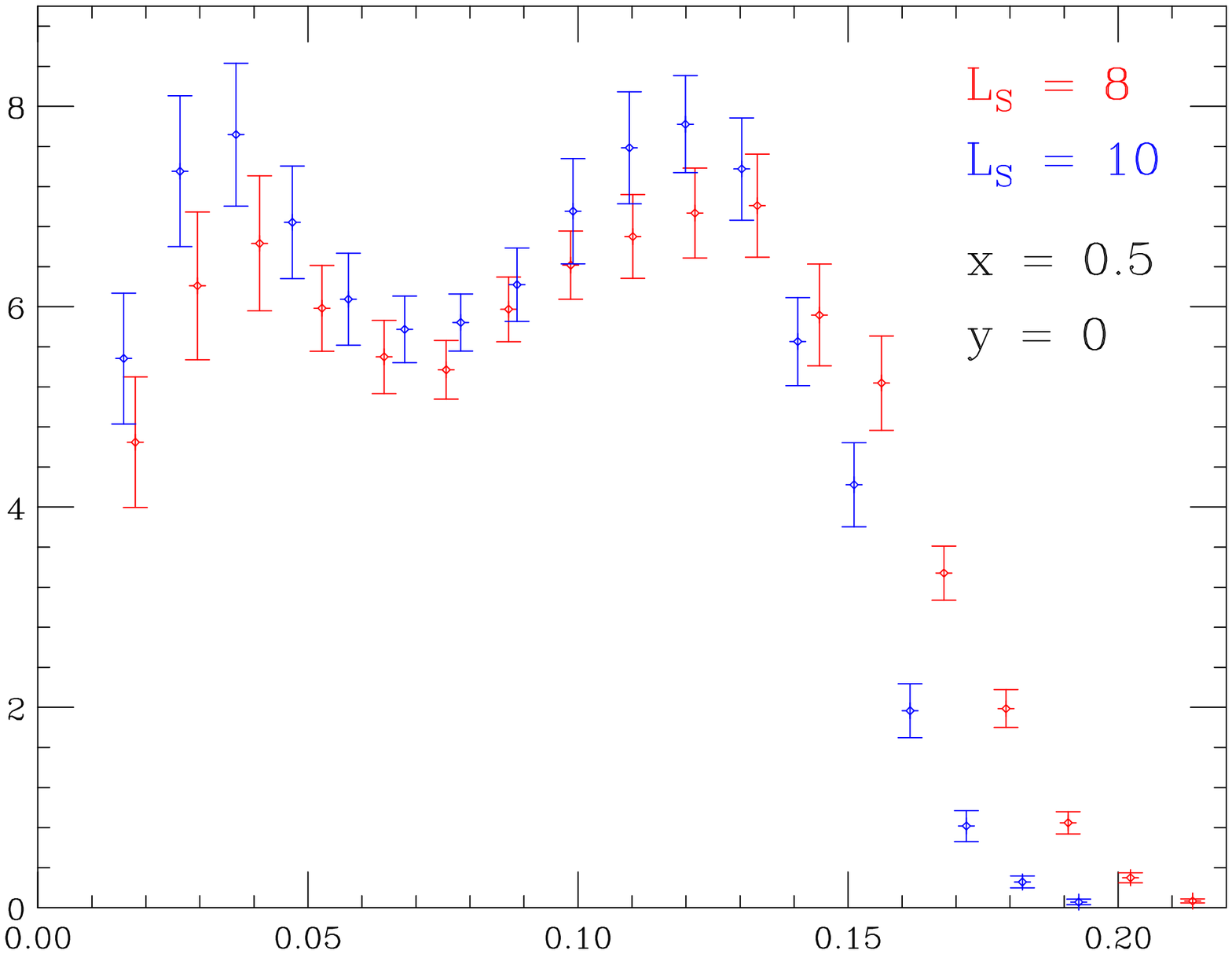,width=.7\textwidth,angle=90}}
\centerline{\epsfig{file=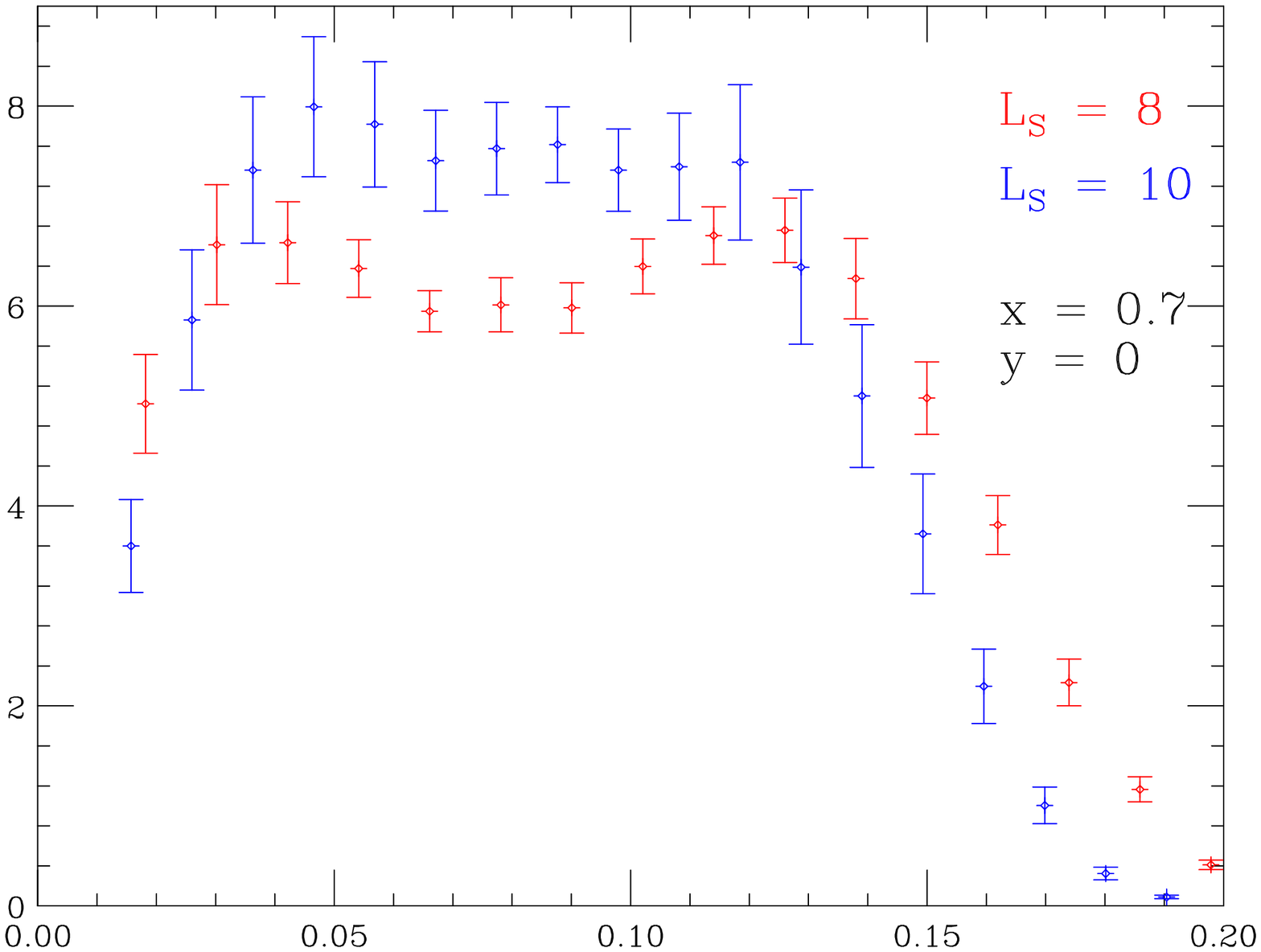,width=.7\textwidth,angle=90}}
\end{minipage}
\caption{Histograms of the modulus of the Polyakov loop in two flavor
QCD.}
\label{fig4}
\end{figure}

\end{document}